\documentclass[fleqn,usenatbib]{mnras}

\usepackage{newtxtext,newtxmath}

\usepackage[T1]{fontenc}

\DeclareRobustCommand{\VAN}[3]{#2}
\let\VANthebibliography\thebibliography
\def\thebibliography{\DeclareRobustCommand{\VAN}[3]{##3}\VANthebibliography}

\usepackage{graphicx}	
\usepackage{amsmath}	
\usepackage{caption}
\usepackage{subcaption}
\usepackage{multirow}

\title[A compact cold molecular outflow in a nearby ULIRG]{ALMA reveals a compact and massive molecular outflow driven by the young AGN in a nearby ULIRG}

\author[L.R. Holden et al.]
{Luke R. Holden$^{1}$,\thanks{E-mail: lholden2@sheffield.ac.uk}
Clive N. Tadhunter$^{1}$,
Anelise Audibert$^{2,3}$,
Tom Oosterloo$^{4,5}$,
Cristina Ramos Almeida$^{2,3}$,
\newauthor
Raffaella Morganti$^{4,5}$,
Miguel Pereira-Santaella$^6$,
and Isabella Lamperti$^7$
\\
$^{1}$Department of Physics $\&$ Astronomy, University of Sheffield, S6 3TG Sheffield, UK. \\
$^{2}$Instituto de Astrofísica de Canarias, Calle Vía Láctea, s/n, 38205 La Laguna, Tenerife, Spain.\\
$^{3}$Departamento de Astrofísica, Universidad de La Laguna, 38206 La Laguna, Tenerife, Spain.\\
$^{4}$ASTRON, the Netherlands Institute for Radio Astronomy, Oude Hoogeveensedijk 4, 7991 PD, Dwingeloo, The Netherlands.\\
$^{5}$Kapteyn Astronomical Institute, University of Groningen, Postbus 800, 9700 AV Groningen, The Netherlands.\\
$^{6}$Instituto de F\'isica Fundamental, CSIC, Calle Serrano 123, 28006 Madrid, Spain \\
$^{7}$Centro de Astrobiología (CAB), CSIC-INTA, Ctra. de Ajalvir Km. 4, 28850 Torrejón de Ardoz, Madrid, Spain.
}

\date{Accepted XXX. Received YYY; in original form ZZZ}

\pubyear{2023}

\begin{document}
\label{firstpage}
\pagerange{\pageref{firstpage}--\pageref{lastpage}}
\maketitle

\begin{abstract}
The ultra luminous infrared galaxy (ULIRG) F13451+1232 is an excellent example of a galaxy merger in the early stages of active galactic nucleus (AGN) activity, a phase in which AGN-driven outflows are expected to be particularly important. However, previous observations have determined that the mass outflow rates of the warm ionised and neutral gas phases in F13451+1232 are relatively modest, and there has been no robust detection of molecular outflows. Using high spatial resolution ALMA CO(1--0) observations, we detect a kiloparsec-scale circumnuclear disk, as well as extended ($r\sim440$\;pc), intermediate-velocity (300\;\textless\;$|v|$\;\textless\;400\;km\;s$^{-1}$) cold molecular gas emission that cannot be explained by rotational disk motions. If interpreted as AGN-driven outflows, the mass outflow rates associated with this intermediate-velocity gas are relatively modest ($\dot{M}_\mathrm{out}=22$--$27$\;M$_\odot$\;yr$^{-1}$); however, we also detect a compact ($r_\mathrm{out}$\;\textless\;120\;pc), high velocity (400\;\textless\;$v$\;\textless\;680\;km\;s$^{-1}$) cold molecular outflow near the primary nucleus of F13451+1232, which carries an order of magnitude more mass ($\dot{M}_\mathrm{out}$$\sim$230\;M$_\odot$\;yr$^{-1}$) than (and several times the kinetic power of) the previously-detected warmer phases. Moreover, the similar spatial scales of this compact outflow and the radio structure indicate that it is likely accelerated by the small-scale ($r$$\sim$130\;pc) AGN jet in the primary nucleus of F13451+1232. Considering the compactness of the nuclear outflow and intermediate-velocity non-rotating gas that we detect, we argue that high spatial-resolution observations are necessary to properly quantify the properties of AGN-driven outflows and their impacts on host galaxies.
\end{abstract}




\section{Introduction}
\label{section: introduction}

Although often required by models of galaxy evolution to explain the observed properties of the galaxy population (e.g. \citealt{Magorrian1998, Silk1998, DiMatteo2005, Dave2019}), considerable uncertainties remain regarding the impact of gas outflows driven by active galactic nuclei (AGN). Principally, this is because outflow masses and kinetic powers are often not precisely determined (see \citealt{Harrison2018}), and their dominant driving mechanisms (radiation pressure e.g. \citealt{Hopkins2010, Meena2023}, or jet-ISM interactions e.g. \citealt{Mukherjee2016, Audibert2023}) in different object types are unclear.

An excellent object to investigate the impact and launching mechanism(s) of AGN-driven outflows is the ultraluminous infrared galaxy (ULIRG; $L_\mathrm{IR}$\;\textgreater\;$10^{12}$\;L$_\mathrm{\odot}$: \citealt{Sanders1996}) F13451+1232 --- also known as 4C12.50 --- which is a merger (\citealt{Gilmore1986, Heckman1986}) in a late pre-coalescence phase (nuclear separation $\sim$4.5\;kpc: \citealt{Tadhunter2018}). ULIRGs such as F13451+1232 are among the most rapidly evolving galaxies in the local universe due to merger-induced gas inflows feeding the central AGN and triggering star formation --- the exact situation modelled in some hydrodynamic simulations of galaxy evolution that include AGN-driven outflows as a feedback mechanism (e.g. \citealt{DiMatteo2005, Hopkins2008}). 

As a luminous radio source (L$_\mathrm{1.4\;GHz} = 1.9\times10^{26}$\;W\;Hz$^{-1}$), F13451+1232 contains radio emission on two distinct scales: small-scale emission within a radius of $\sim$130\;pc of the primary nucleus along position angle PA=151$^\circ$ \citep{Stanghellini1997, Lister2003, Morganti2013_4c1250}, and larger-scale, diffuse emission extending to a radial distance of $\sim$77\;kpc from the primary nucleus along PA$\sim$180$^\circ$ \citep{Stanghellini2005}. It has been proposed that this radio structure could be the result of two epochs of jet activity \citep{Stanghellini2005}, with the smaller-scale radio structure comprising a young, recently-restarted jet, and the larger-scale emission originating from a previous jet cycle. 

Because of the properties of its small-scale radio emission, F13451+1232 is also classified as a gigahertz-peaked-spectrum (GPS) source (see \citealt{ODea2021} for a review); such sources are associated with powerful jet-ISM interactions that may accelerate gas outflows in multiple phases \citep{Mukherjee2016, Holt2006, Santoro2018, Santoro2020, Kukreti2023}. In addition, given the object's high optical emission-line luminosity (L$_\mathrm{[OIII]}=1.45\times10^{43}$\;erg\;s$^{-1}$: \citealt{Rose2018}) --- leading to its additional classification as a type-2 quasar (QSO2)\footnote{F13451+1232 is included in the QUADROS sample of ULIRGs \citep{Rose2018}, and the QSOFEED sample of type-2 quasars \citep{RamosAlmeida2022}.} --- it might also be expected to accelerate outflows via radiation-pressure-driven winds.

Despite the fact that outflows in F13451+1232 have been confirmed in the coronal \citep{VillarMartin2023}, warm ionised \citep{Holt2003, Holt2011, Rose2018, VillarMartin2023}, and neutral atomic (HI + NaID: \citealt{Morganti2005, Rupke2005, Morganti2013_4c1250}) gas phases, there has so far not been a robust detection of molecular outflows in this object (\citealt{Fotopoulou2019, Lamperti2022}; see discussion in \citealt{VillarMartin2023}). In nearby AGN where molecular outflows have been detected, it is often found that they contain more mass and kinetic power than the warmer phases (e.g. \citealt{Fiore2017, RamosAlmeida2019, Holden2023, HoldenTadhunter2023, Speranza2023}). In this context, it is important to note that the warm ionised and neutral atomic outflows in F13451+1232 have relatively modest mass outflow rates ($\sim$6--$12$\;M$_\odot$\;yr$^{-1}$: \citealt{Morganti2013_4c1250, Rose2018}); therefore, if molecular outflows are present, they could potentially carry enough mass to change the interpretation AGN feedback in this important object. 

A cold molecular outflow in a similar source, PKS 1549-79, was found to be compact ($r$\;\textless\;120\;pc: \citealt{Oosterloo2019}), as were the previously-detected neutral HI ($r_\mathrm{HI}$\textless\;100\;pc: \citealt{Morganti2013_4c1250}) and warm ionised ($r_\mathrm{[OIII]}\sim$69\;pc: \citealt{Tadhunter2018}) outflows in F13451+1232. Therefore, here we present high-spatial resolution ($0.113\times0.091$\;arcsecond or $247\times119$\;pc beam-size), high sensitivity ALMA observations of the inner few kiloparsecs of the primary nucleus of F13451+1232 to search for (and characterise) cold molecular outflows in CO(1--0) emission.

In Section \ref{section: observations_and_data_reduction}, we describe the ALMA CO(1--0) observations and our data reduction process; in Section \ref{section: analysis_and_results} we detail our analysis of the cold molecular gas in the central kiloparsecs of F13451+1232; in Section \ref{section: discussion} we discuss our findings and their wider implications, and in Section \ref{section: conclusions} we give our conclusions.

Throughout this work, we assume a cosmology of $H_0=70$\;km\;s$^{-1}$\;Mpc$^{-1}$, $\Omega_\mathrm{m}=0.3$, and $\Omega_\mathrm{\lambda}=0.7$. This corresponds\footnote{Calculated using Ned Wright's Javascript Cosmology Calculator \citep{Wright2006}.} to an arcsecond-to-kpc spatial conversion factor of 2.189\;kpc/arcsec and a luminosity distance of $D_\mathrm{L}=570$\;Mpc at the redshift of F13451+1232 ($z=0.121680$: \citealt{Lamperti2022}).

\section{Observations and data reduction}
\label{section: observations_and_data_reduction}

F13451+1232 was observed in Band 3 of the Atacama Large Millimeter/submillimeter Array (ALMA) on two nights using two different configurations of the 12\;m array (Project Code 2019.1.01757.S): the array configuration used on the 10th September 2021 (baselines $=180$--16200\;m) resulted in a beam size of $0.083\times0.069$\;arcseconds, while the configuration used on the 28th September 2021  (baselines $=70$--14400\;m) resulted in a beam size 0.166$\times$0.123\;arcseconds. The total time on-source was 5588\;seconds ($\sim$1.5\;hours). Four spectral windows were used for the observations: one centred on 102.790\;GHz with a bandwidth of 1.875\;GHz (covering the CO(1--0) line), and the remaining three centred on 104.581\;GHz, 90.977\;GHz, and 92.477\;GHz with bandwidths of 2.000\;GHz (covering the continuum). J1337-1257 and J1353+1435 were observed to provide bandpass/flux and phase calibration, respectively; phase self-calibration was performed using the \textsc{Miriad} data reduction package \citep{Sault1995}.

The CO(1--0) transition was targeted for two reasons. First, the ALMA observing band that covers the CO(1--0) line is wider than those that cover higher-order transitions such as CO(2--1) or CO(3--2), and therefore facilitates more accurate continuum subtraction. Second, the CO(1--0) transition is less susceptible to gas excitation conditions than the higher-order CO transitions, and hence may be a more reliable probe of molecular gas masses in ULIRGs.

In order to check for instrumental errors such as bandpass and spectral calibration issues, we produced three CO(1-0) line datacubes: two produced using the visibilities of the two different array configurations separately, and one produced by combining the visibilities from both observations (with a final beam size of $0.113\times0.091$\;arcseconds, PA$=-51.8^\circ$). To optimise the column density sensitivity, and to ensure that the dirty beam was not significantly non-Gaussian, each CO(1--0) line cube was made using Briggs weighting with \mbox{robust $=1$}. When inspecting these cubes individually, we did not find any indications of bandpass or spectral calibration issues.

The data products produced by the automated ALMA pipeline are potentially affected by issues related to continuum subtraction, which are a consequence of the bright and steep continuum emission produced by the primary nucleus of F13451+1232. Therefore, for all three datacubes, we started with the respective visibilities as calibrated by the ALMA pipeline and applied a standard self-calibration using continuum images made from the line-free channels. We then created the line datacubes, and removed the continuum in each by subtracting a linear function that was fitted to the continuum on either side of (but excluding) the CO(1--0) emission. A continuum image was also generated from the combined visibilities of both observations using the continuum spectral windows. We used least-squares fitting with the \textsc{AstroPy} \citep{AstropyCollaboration2013, AstropyCollaboration2018} \textsc{Python} module to fit a two-dimensional Gaussian to the point source in this continuum image, and took the centroid position (13:47:33.36 +12 17 24.23) to be the location of the nucleus. The continuum image was made using uniform weighting and has a resolution of $0.049\times0.033$\;arcseconds (PA$=-46.8^\circ$).

To ensure the highest sensitivity and signal-to-noise, we henceforth only consider the combined CO(1--0) datacube in our main analysis, which was produced using the combined visibilities of both observations; we discuss results from the cubes produced using the two observations separately in Appendix \ref{appendix: beam_sizes}. The combined cube has a root-mean-square (RMS) noise of $0.148$\;mJy\;beam$^{-1}$ for a velocity resolution of 28.8\;km\;s$^{-1}$.

\section{Analysis and results}
\label{section: analysis_and_results}

\subsection{Moment maps}
\label{section: analysis_and_results: moment_maps}

\begin{figure*}
    \centering
    \includegraphics[width=1\linewidth]{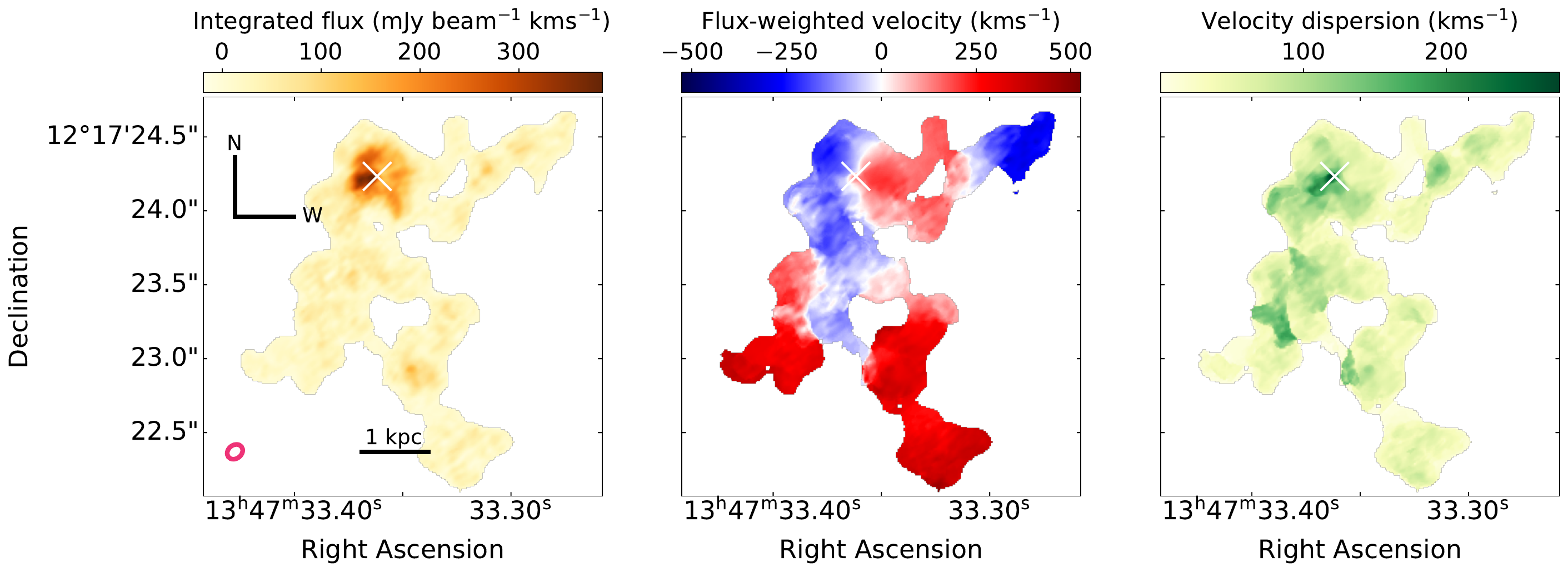}
    \caption{Integrated flux (moment 0), flux-weighted velocity (moment 1), and velocity dispersion (moment 2) maps for the central $\sim$6\;kiloparsecs of the primary nucleus of F13451+1232. The white cross marks the continuum centre, which we take to be the position of the AGN nucleus. The beam size of the observations is shown as a pink ellipse in the bottom left corner, and a 1\;kpc scale bar is shown in black.}
    \label{fig: large_fov_moment_maps}
\end{figure*}

\begin{figure*}
    \centering
    \includegraphics[width=1\linewidth]{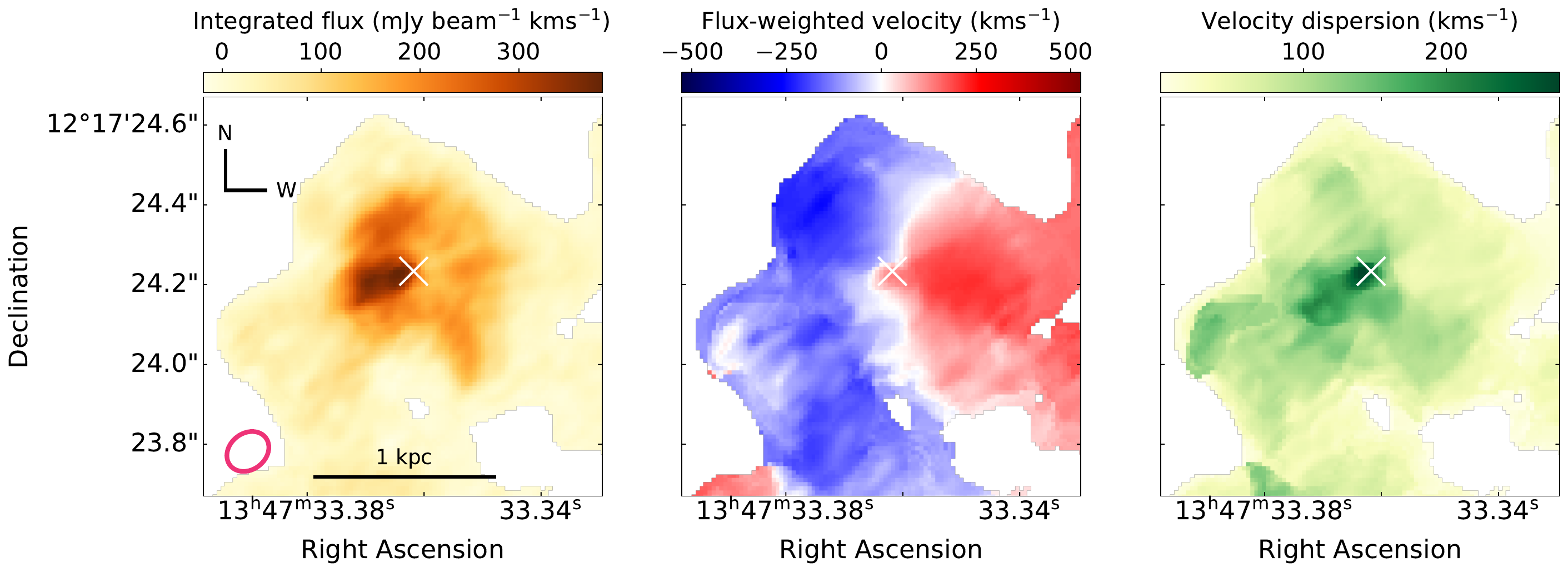}
    \caption{Integrated flux (moment 0), flux-weighted velocity (moment 1), and velocity dispersion (moment 2) maps for the inner kiloparsec region of the primary nucleus of F13451+1232; the marker and symbol scheme is the same as in Figure \ref{fig: large_fov_moment_maps}.}
    \label{fig: moment_maps}
\end{figure*}

To investigate the flux and kinematic structures of the CO(1--0) emission around the primary nucleus of F13451+1232, we first created moment maps of the combined datacube using the \textsc{SoFiA-2} source-finding pipeline \citep{Serra2015, Serra2021}. First, \textsc{SoFiA-2} applied different combinations of spatial and velocity smoothing to the datacube: Gaussian filters with FWHMs of 0, 3, 6, 12, and 15 pixels were used for the spatial smoothing, and boxcar filters of widths 0, 3, 7, and 15 channels were used for the velocity smoothing. For each combination of spatial and velocity smoothing, emission above $4\sigma_\mathrm{rms}$ was added to a total mask. This mask was then applied to the original, non-smoothed datacube, from which the maps in Figure \ref{fig: large_fov_moment_maps} and \ref{fig: moment_maps} were produced. 

In the $6\times6$\;kpc moment maps (Figure \ref{fig: large_fov_moment_maps}), cold molecular gas is detected to radial distances of $\sim$2.5\;arcseconds ($\sim$5.5\;kpc) south of the primary nucleus, and $\sim$1\;arcsecond ($\sim$2\;kpc) to the west. This large-scale ($r$\;\textgreater\;1\;kpc) emission, which has velocities in redshift and blueshift of up to $v$$\sim$|500|\;km\;s$^{-1}$, is indicative of gas that has been disturbed by the merger.

Within the central kiloparsec of the primary nucleus (Figure \ref{fig: moment_maps}), there is clear evidence for a disk with radius $r_\mathrm{disk}$$\sim$0.5\;kpc centred on the nucleus, blueshifted to the east and redshifted to the west, with a projected maximum flux-weighted velocity of 200--250\;km\;s$^{-1}$. In addition, there is an extended region (up to $r$$\sim$0.3\;arcseconds or 660\;pc from the nucleus along PA$=120^\circ$) of emission with higher velocity dispersions ($\sigma$\;\textgreater\;150\;km\;s$^{-1}$) than that of the disk. Furthermore, a particularly prominent redshifted component with a velocity dispersion of $\sim$250\;km\;s$^{-1}$ is seen within $\sim$0.1\;arcseconds ($\sim$220\;pc) of the continuum centre.

\subsection{Properties of the kpc-scale disk}
\label{section: analysis_and_results: disk}

To separate non-gravitational kinematics from those expected from a rotating gas disk, we modelled the disk using the \textsc{3DFIT} task from the \textsc{BBarolo} tool \citep{DiTeodoro2015}; the fitting process and results are detailed in Appendix \ref{appendix: disk_modelling}. From the \textsc{BBarolo} model, we find that the disk has an average inclination of $i\approx54^\circ$ (where $i=0^\circ$ corresponds to the disk being face-on) and that its major axis lies along PA$\approx248^\circ$. The deprojected rotational velocity increases from $v_\mathrm{rot}=246$\;km\;s$^{-1}$ to $v_\mathrm{rot}=307$\;km\;s$^{-1}$ over the radius range 28\;pc\;\textless\;$r_\mathrm{disk}$\;\textless\;560\;pc. We highlight that the purpose of this modelling was solely to give a reasonable estimate of the maximum rotational velocity of the molecular gas, which was needed to robustly identify non-rotational motions in our analysis. Considering this, it is notable that the high velocity-dispersion emission ($\sigma$\;\textgreater\;150\;km\;s$^{-1}$) seen near the disk in the moment maps (Figure \ref{fig: moment_maps}) cannot be accounted for by the \textsc{BBarolo} model (see Figure \ref{fig: bbarolo_model}), and thus cannot be explained as being part of the rotating gas disk.

\subsection{Kinematics of the non-rotating gas}
\label{section: analysis_and_results: outflow_kinematics}

\begin{figure*}
    \centering
    \begin{subfigure}[b]{0.48\textwidth}
        \centering
        \includegraphics[width=\textwidth]{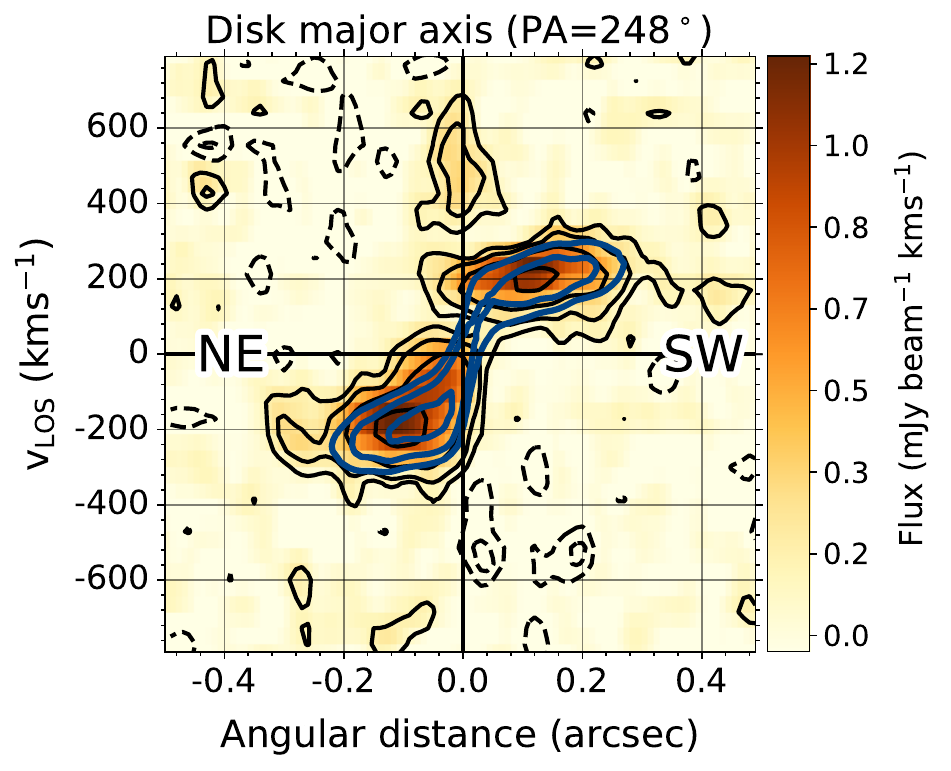}
    \end{subfigure}
    \hspace{1em}
    \begin{subfigure}[b]{0.48\textwidth}
        \centering
        \includegraphics[width=\textwidth]{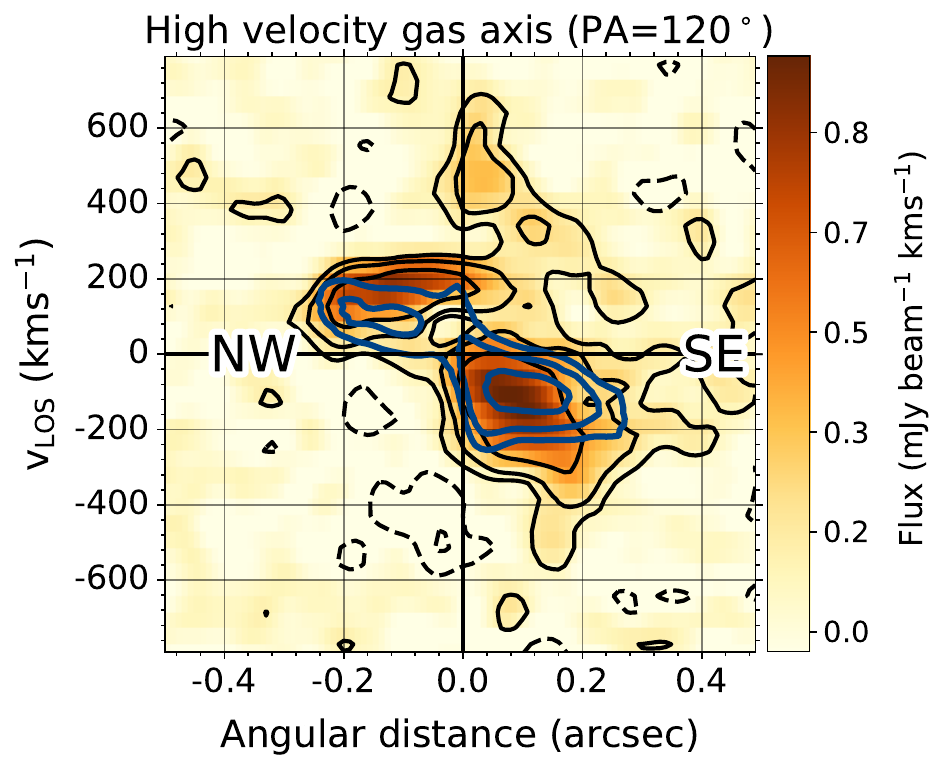}
    \end{subfigure}
    \caption{Position-velocity (PV) diagrams for $1.00\times0.05$\;arcsecond slices with PA=248$^\circ$ (left; the major axis of the disk) and PA=120$^\circ$ (right; the direction of the high velocity dispersion emission seen in Figure \ref{fig: moment_maps}), centred on the continuum centre. CO(1--0) emission is shown in yellow/orange, with black contours at the $-3\sigma_\mathrm{rms}$, $-1.5\sigma_\mathrm{rms}$, 1.5$\sigma_\mathrm{rms}$, 3$\sigma_\mathrm{rms}$, 6$\sigma_\mathrm{rms}$, and 12$\sigma_\mathrm{rms}$ levels (dashed: negative; solid: positive); solid blue contours show the flux from the \textsc{BBarolo} disk model at the same $\sigma_\mathrm{rms}$ levels. Extended, intermediate-velocity emission is seen in both redshift and blueshift in the PA=120$^\circ$ diagram, and a compact, higher-velocity feature (400\;\textless\;$v$\;\textless\;680\;km\;s$^{-1}$) centred approximately on the nucleus (0.0\;arcsec) can clearly be seen in both diagrams.}
    \label{fig: pv_diagrams}
\end{figure*}

To investigate the non-rotational kinematics further, we binned the velocity channels of the original cube by a factor of three and applied Hanning smoothing using the \textsc{CASA} software suite \citep{Bean2022} to improve the signal-to-noise ratio of the emission. This resulted in a cube of velocity resolution of 86.4\;km\;s$^{-1}$, with a RMS noise of $0.083$\;mJy\;beam$^{-1}$. Using the \textsc{pvextractor Python} module\footnote{\url{https://pvextractor.readthedocs.io/en/latest/}} we extracted $1.00\times0.05$\;arcsecond slices, centred on the nucleus, from both this datacube and the \textsc{BBarolo} disk-model datacube along the major axis of the disk (PA=248$^\circ$) and in the direction of the high-velocity dispersion gas near the nucleus (PA=120$^\circ$). From these slices, we produced position-velocity (PV) diagrams, which we present in Figure \ref{fig: pv_diagrams}. In the PV diagrams, there is clear evidence for non-rotational kinematics, with detected velocities being well above the projected maximum rotational disk velocity given by the \textsc{BBarolo} model. There is intermediate velocity emission (300\;\textless\;$|v|$\;\textless\;400\;km\;s$^{-1}$) seen in both blueshift and redshift along PA$=120^\circ$, in the ranges 0.1\;\textless\;$r$\;\textless\;0.3\;arcseconds and 0.0\;\textless\;$r$\;\textless\;0.2\;arcseconds, respectively, to the southeast (SE) of the nucleus. Moreover, the compact ($r$\;\textless\;0.1\;arcseconds) redshifted feature visible in the moment maps (Figure \ref{fig: moment_maps}) can be seen in the PV diagrams between 400\;\textless\;$v$\;\textless\;680\;km\;s$^{-1}$, located close to the nucleus.

\subsection{Comparison to the small-scale radio structure}
\label{section: analysis_and_results: radio_structure}

\begin{figure}
    \centering
    \includegraphics[width=\linewidth]{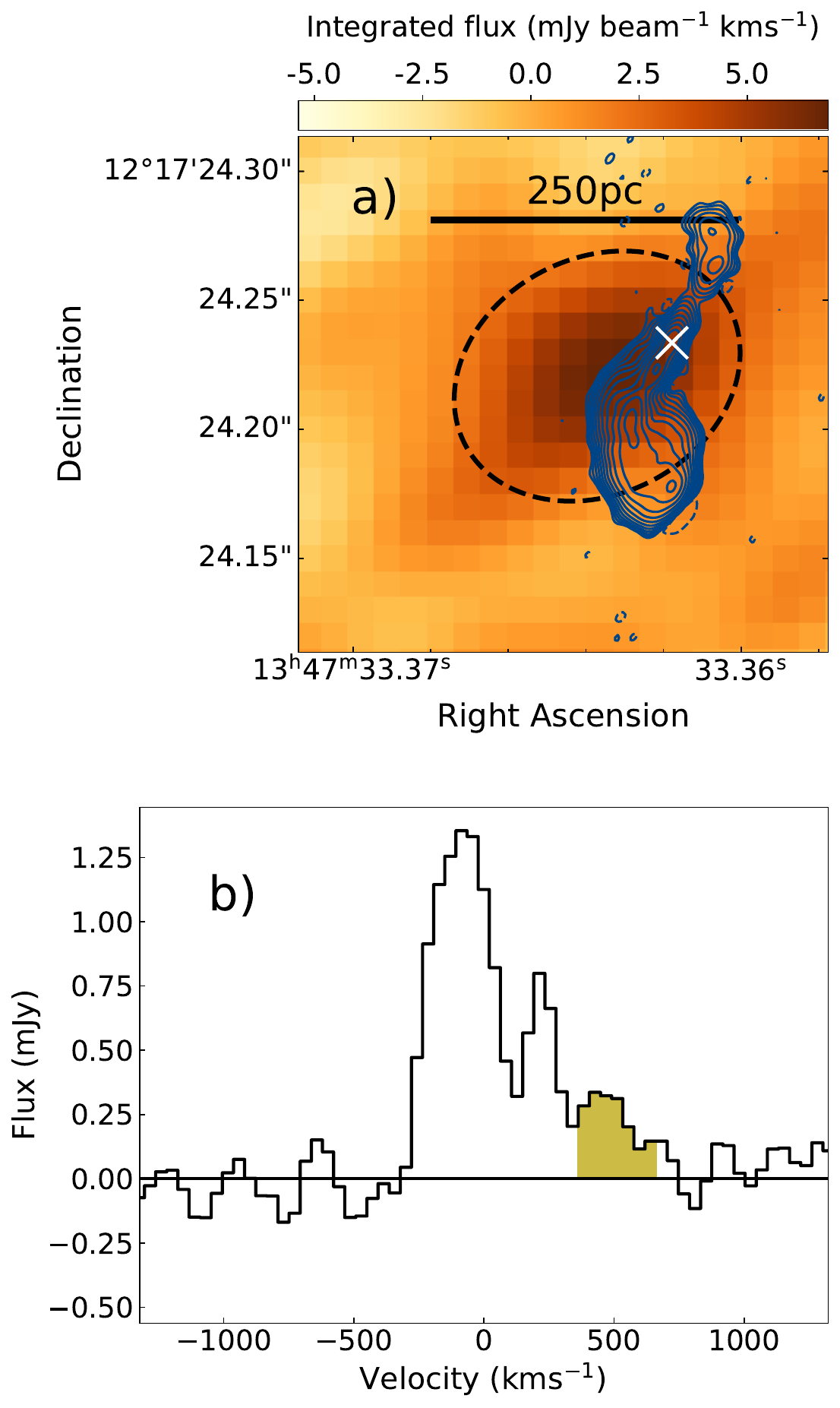}
    \caption{\textbf{a)} Integrated flux map (orange--brown) of the velocity channels between 400\;\textless\;$v$\;\textless\;680\;km\;s$^{-1}$. The dashed black ellipse shows the aperture extracted from the datacube to derive the mass of the outflow; the corresponding line profile is shown in Figure \ref{fig: outflow_moment_map}b. The small-scale radio structure, as seen in VLBI 1266\;MHz continuum imaging \citep{Morganti2013_4c1250}, is shown as dark blue solid contours at the $0.3\times$(-1, 1, 2, 4, 8, 16, 32, 64, 128, 256, 512, 1024) mJy/beam levels, whereas the white cross marks the position of the continuum centre seen in our observations (Section \ref{section: observations_and_data_reduction}). \textbf{b)} CO(1--0) line profile extracted from the aperture that covers the outflow; the line profile shown was extracted from the 84\;km\;s$^{-1}$ velocity resolution combined datacube. The part of the line profile that we take to represent in-/out-flowing gas (400\;\textless\;$v$\;\textless\;680\;km\;s$^{-1}$) is shaded in yellow.}
    \label{fig: outflow_moment_map}
\end{figure}

Due to the velocities at which the compact, high-velocity nuclear feature is detected, it is unambiguously distinct from the rotating gas disk: it may be interpreted as either an inflow moving towards the nucleus on the observer's side of the disk, or an outflow originating from the nucleus on the far side of the disk. Spatial comparison of the high-velocity flow to the small-scale radio structure may provide an indication of the nature of this flowing gas and, if it is an outflow, the driving mechanism. In Figure \ref{fig: outflow_moment_map}a, we show the flux map integrated over the range 400\;\textless\;$v$\;\textless\;680\;km\;s$^{-1}$ --- in which we detect peak emission above the 6$\sigma_\mathrm{rms}$ level --- along with VLBI 1266\;MHz continuum imaging (from global VLBI experiment GM62B; data presented and detailed in \citealt{Morganti2013_4c1250}) of the central $0.20\times0.20$\;arcseconds ($440\times440$\;parsecs). The beam size of the VLBI observations was $8.01\times4.89$\;milliarcseconds (PA$=-20.2^\circ$). Due to the small uncertainties in the ALMA pointing, the VLBA image was offset from our continuum centre by $\sim$0.1\;arcseconds. We accounted for this pointing error by aligning the compact core seen in the VLBA image with the continuum centre, as it is expected that the same source is being imaged at different spatial resolutions. 

By fitting a 2D Gaussian profile to the integrated high-velocity CO(1--0) emission in Figure \ref{fig: outflow_moment_map}a using the \textsc{emcee} Python module \citep{FormanMackey2013}, we find it to be offset to the southeast (SE) of the continuum centroid by (3.2$\pm$0.2)$\times10^{-2}$\;arcseconds (70$\pm$4\;pc) along PA$\sim$155$^{\circ}$. We highlight that this is within the spatial extent of, and has a similar PA to, the small-scale radio structure ($r\sim0.07$\;arcsec; $\mathrm{PA}=151^\circ$)

\subsection{Channel maps}
\label{section: analysis_and_results: channel_maps}

\begin{figure}
    \centering
    \includegraphics[width=1\linewidth]{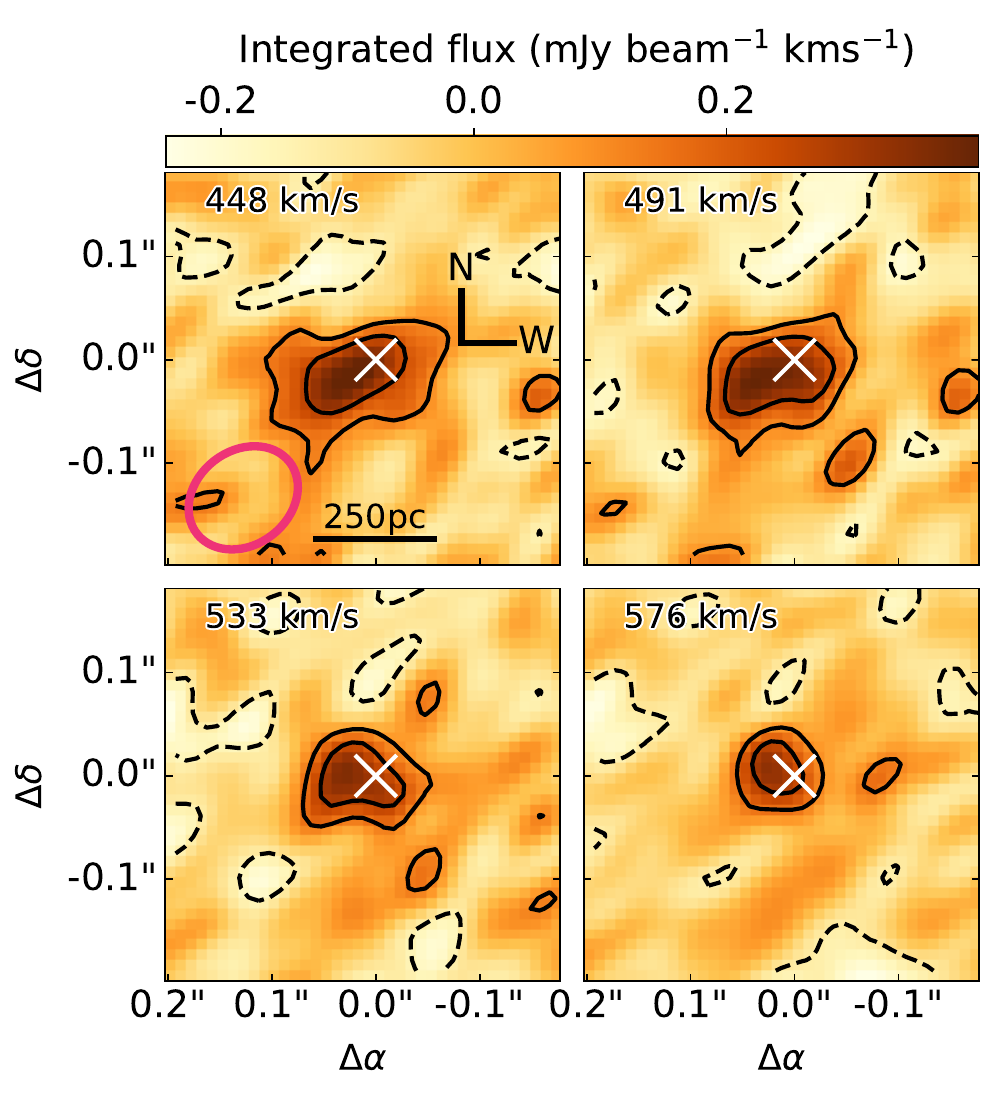}
    \caption{Channel maps of the central $\sim$1\;kpc of the primary nucleus of F13451+1232 (centred on the nucleus: Section \ref{section: observations_and_data_reduction}) between 448\;\textless\;$v$\;\textless\;576\;km\;s$^{-1}$, taken from the velocity-binned and smoothed cube with channel widths of 43.7\;km\;s$^{-1}$ and a velocity resolution of 86.4\;km\;s$^{-1}$. Black contours are shown at the $-3\sigma_\mathrm{rms}$, $-1.5\sigma_\mathrm{rms}$, 1.5$\sigma_\mathrm{rms}$, 3$\sigma_\mathrm{rms}$ levels (dashed: negative; solid: positive), where $\sigma_\mathrm{rms}$ is the RMS of the binned-and-smoothed cube. The continuum centre is marked as a white cross in each channel; the beam size is shown as a pink ellipse in the 448\;km\;s$^{-1}$ channel map (top left panel),}
    \label{fig: channel_maps}
\end{figure}

To determine the level at which the high-velocity, nuclear \mbox{CO(1--0)} emission is detected in different velocity channels, we created channel maps of the central kiloparsec of the primary nucleus from the 86.4\;km\;s$^{-1}$ velocity resolution cube. The high-velocity (400\;\textless\;$v$\;\textless\;600\;km\;s$^{-1}$) channel maps --- presented in Figure \ref{fig: channel_maps} --- show emission above the 3$\sigma_\mathrm{rms}$ level (and up to 4$\sigma_\mathrm{rms}$) in all four channels on scales similar to that of the beam size (0.11\;arcseconds or 240\;pc), indicating that the feature is spatially-unresolved in the combined datacube. 


\subsection{Potential effects of continuum subtraction}
\label{section: analysis_and_results: potential_effects_of_continuum_subtraction}

Due to the bright, steep-spectrum continuum around the CO(1--0) line --- originating from primary nucleus of F13451+1232 (as seen in the cores of other CSS/GPS sources, e.g. \citealt{Oosterloo2019}) --- it is possible that inaccuracies in the continuum subtraction process have contributed flux to the compact, redshifted, high-velocity feature that is detected at the position of the nucleus. However, we argue that this feature represents real CO(1--0) emission, for several reasons. Firstly, the flux is detected above the $6\sigma$ level in the flux map integrated over the velocity range 400\;\textless\;$v$\;\textless\;680\;km\;s$^{-1}$ (Figure \ref{fig: outflow_moment_map}a), and in multiple velocity channels at the 3--4$\sigma_\mathrm{rms}$ levels (Figure \ref{fig: channel_maps}). Secondly, it has a significantly higher flux than the negative or positive residual features detected in the integrated spectrum (Figure \ref{fig: outflow_moment_map}b). Finally, and most importantly, the feature is detected at the same velocity --- and at higher flux than any residual features --- in both of the two observations that were used to produce the combined datacube, despite their differing spatial resolutions (see Appendix \ref{appendix: beam_sizes}). While we cannot entirely rule out some contribution from instrumental effects, we conclude that this feature is real.

It is notable that previous CO(2--1) observations of the primary nucleus of F13451+1232 --- that were of similar sensitivity to our CO(1--0) observations --- did not detect evidence for nuclear outflows \citep{Lamperti2022}. However, in those observations, the other spectral windows were also used when fitting a linear slope to the continuum because of the narrower observing band (a factor of two narrower in velocity) of CO(2--1) observations relative to CO(1--0) --- this may have led to over-subtraction of the high-velocity outflow component. Therefore, it is possible that outflow emission in CO(2--1) was not detected due to uncertainties in the continuum subtraction process.

\subsection{Outflow energetics}
\label{section: analysis_and_results: outflow_energetics}

In order to provide an estimate of the mass inflow/outflow rates of the high-velocity, spatially-unresolved redshifted feature located at the position of the nucleus, we extracted an elliptical aperture matching the beam size ($0.113\times0.091$\;arcsec, PA$=-51.8^\circ$; shown in Figure \ref{fig: outflow_moment_map}a) from the datacube. We then summed all pixels within the aperture to create a line profile, which we present in Figure \ref{fig: outflow_moment_map}b, and selected emission between 400\;\textless\;$v$\;\textless\;680\;km\;s$^{-1}$ to isolate the non-rotating gas. 

For the extended, intermediate-velocity (300\;\textless\;$|v|$\;\textless\;400\;km\;s$^{-1}$) emission seen in blueshift and redshift, we extracted two sub-apertures from the $1.00\times0.05$\;arcsecond slice along PA$=120^\circ$ (used to create the right PV diagram in Figure \ref{fig: pv_diagrams}): one covering radii 0.0\;\textless\;$r$\;\textless\;0.2\;arcseconds (for the extended redshifted emission), and the other covering the range 0.1\;\textless\;$r$\;\textless\;0.3\;arcseconds (for the extended blueshifted emission). We then selected emission between 300\;\textless\;$v$\;\textless\;400\;km\;s$^{-1}$ and $-400$\;\textless\;$v$\;\textless$-300$\;km\;s$^{-1}$ respectively in these slices to isolate the non-rotating gas.

For every channel within the range of the selected velocities in the extracted apertures, we calculated the CO(1--0) fluxes ($S_\mathrm{CO}\Delta V$) and converted them to luminosities ($L^\prime_\mathrm{CO}$) using the relation presented by \citet{Solomon2005}:
\begin{equation}
    L^\prime_\mathrm{CO} = 3.25\times10^7 S_\mathrm{CO}\Delta V {\nu}^{-2}_\mathrm{obs}D^2_\mathrm{L}(1+z)^{-3},
    \label{eq: co_luminosity}
\end{equation}
where ${\nu}_\mathrm{obs}$ is the observed frequency of the CO(1--0) line (102.789\;GHz) and $D_\mathrm{L}$ is the luminosity distance to F13451+1232. We then summed the CO(1--0) luminosities across all selected velocity channels, resulting in total luminosities for each of the two intermediate-velocity features and the high-velocity feature. Assuming a CO-to-H$_\mathrm{2}$ conversion factor of $\alpha_\mathrm{CO}=0.8$\;M$_\odot$\;(K\;km\;s$^{-1}$\;pc$^2$)$^{-1}$ (typical for ULIRGs: \citealt{Downes1998}), we converted these total luminosities into molecular gas masses. We also applied this method to the rotating kpc-scale CO disk, for which we estimate a total mass of $1.1\times10^{10}$\;M$_\odot$. 

The corresponding mass flow rates of the non-rotating molecular gas were estimated using
\begin{equation}
    \dot{M}_\mathrm{flow} = \frac{M_\mathrm{flow}v_\mathrm{w}}{\Delta R} \cdot \mathrm{tan\theta},
\end{equation}
where $v_\mathrm{w}$ is the flux-weighted velocity, $\Delta R$ is the flow radius, and $\theta$ is the angle of the outflow relative to our line of sight. The $\mathrm{tan\theta}$ term is required to correct for projection effects in both radius and velocity. To estimate this projection factor, we assumed a solid-angle-weighted mean inclination for random orientations of the line of sight to the outflow direction, giving $\theta=57.3^\circ$. Finally, we estimated kinetic powers with
\begin{equation}
    \dot{E}_\mathrm{kin}=\frac{1}{2}\dot{M}_\mathrm{flow}\Bigl(\frac{v_\mathrm{w}}{\mathrm{cos\theta}}\Bigl)^2.
    \label{eq: kinetic_power}
\end{equation}
We present the resulting CO(1--0) luminosities, masses, flux-weighted velocities, flow radii, mass flow rates, kinetic powers, and coupling efficiencies ($\epsilon_\mathrm{f}=\dot{E}_\mathrm{kin}$/$L_\mathrm{bol}$) in Table \ref{tab: outflow_properties}. For the high-velocity nuclear gas, we derive a mass flow rate of $\dot{M}_\mathrm{out}\sim$230\;M$_\odot$\;yr$^{-1}$, which corresponds to a kinetic power that is $\sim$1.4\;per\;cent of the bolometric luminosity of the AGN ($L_\mathrm{bol}=4.8\times10^{45}$\;erg\;s$^{-1}$: \citealt{Rose2018}). In contrast, we find much lower mass flow rates ($\dot{M}_\mathrm{out}=22$--$27$\;M$_\odot$\;yr$^{-1}$) and kinetic powers ((6.0--7.1)$\times10^{-2}$\;per\;cent of $L_\mathrm{bol}$) for the extended, intermediate-velocity emission.

\begin{table*}
    \centering
    \begin{tabular}{l|c|c|c|c|c|c|c|c}
        \multirow{2}{*}{Component} & $L^\prime_\mathrm{CO}$  & $M_\mathrm{flow}$  & $v_\mathrm{w}$  & ${\Delta}R$ $^1$ & $\dot{M}_\mathrm{flow, proj}$ & $\dot{M}_\mathrm{flow}$  & $\dot{E}_\mathrm{kin}$  & $\epsilon_\mathrm{f}$ $^2$ \\
          & [K\;km\;s$^{-1}$\;pc$^{2}$] & [M$_\odot$] & [km\;s$^{-1}$] & [arcsec] &  [M$_\odot$\;yr$^{-1}$] & [M$_\odot$\;yr$^{-1}$] & [erg\;s$^{-1}$] & [per\;cent] \\
        \hline
        Extended blueshifted & $2.7\times10^7$ &  $2.2\times10^7$  & $-339$ & 0.2 & 17 & 27 & $3.4\times10^{42}$ & $7.1\times10^{-2 }$  \\
        Extended redshifted &  $2.2\times10^7$  & $1.8\times10^7$ & 350 & 0.2 & 14 & 22 &  $2.9\times10^{42}$ & $6.0\times10^{-2}$ \\
        Nuclear redshifted & $4.2\times10^7$ & $3.4\times10^7$ & 514 & 0.0545 & 148 & 230 & $6.6\times10^{43}$ & 1.4
    \end{tabular} \\
    \centering
    $^1$For the extended emission, we take ${\Delta}R$ to be the radial distance over which the extended emission is seen; for the spatially unresolved, high-velocity redshift feature, we take ${\Delta}R$ to be half of the beam major axis. \\
    $^2$Calculated assuming $L_\mathrm{bol}=4.8\times10^{45}$\;erg\;s$^{-1}$ \citep{Rose2018}.
    \caption{CO(1--0) luminosities, masses, projected flux-weighted outflow velocities, aperture radii, projected mass outflow rates ($\dot{M}_\mathrm{flow, proj}$), deprojected mass outflow rates ($\dot{M}_\mathrm{flow}$), and deprojected energetics for the extended emission in both blueshift and redshift, and the nuclear redshifted feature seen in the PV diagrams (Figure \ref{fig: pv_diagrams}). When deprojecting, we assumed $\theta=57.3^\circ$ (see Section \ref{section: analysis_and_results: outflow_energetics}).}
    \label{tab: outflow_properties}
\end{table*}

\section{Discussion}
\label{section: discussion}

In addition to a kpc-scale disk, we have detected non-rotating cold molecular gas in CO(1--0) emission near the primary nucleus of F13451+1232. This non-rotating gas consists of spatially-extended (0.0\;\textless\;$r$\;\textless\;0.3\;arcseconds; 0\;\textless\;$r$\;\textless\;660\;pc), intermediate-velocity (300\;\textless\;$|v|$\;\textless\;400\;km\;s$^{-1}$) components seen in both blueshift and redshift, and a compact ($r$\;\textless\;0.06\;arcseconds; $r$\;\textless\;120\;pc), spatially-unresolved, high-velocity (400\;\textless\;$v$\;\textless\;680\;km\;s$^{-1}$) nuclear component seen in redshift.

\subsection{Interpreting the non-rotating gas as inflowing}
\label{section: discussion: inflows}

Given that the AGN in F13451+1232 may have recently restarted and thus be young \citep{Stanghellini2005, Morganti2013_4c1250}, if the non-rotating gas seen in intermediate and high-velocity CO(1--0) emission is inflowing, it is possible that we are directly observing the gas flows responsible for its triggering. The total gas flow rates (27--230\;M$_\odot$\;yr$^{-1}$: Table \ref{tab: outflow_properties}) are orders of magnitude higher than those needed to sustain the AGN (0.8\;M$_\odot$yr$^{-1}$, assuming $\eta=0.1$ and $L_\mathrm{bol}=4.8\times10^{45}$\;erg\;s$^{-1}$: \citealt{Rose2018}). Therefore, this is a plausible feeding mechanism even if only a small fraction of the inflowing gas is accreted onto the central supermassive black hole (SMBH). However, we argue below that the compact, high-velocity component (and potentially also the intermediate velocity emission) is in reality outflowing, rather than inflowing.

\subsection{The spatially-extended, intermediate-velocity gas}
\label{section: discussion: intermediate_velocity_gas}

The nature of the intermediate velocity emission (300\;\textless\;$|v|$\;\textless\;400\;km\;s$^{-1}$) that is seen in both redshift and blueshift in the PV diagram along PA$=120^\circ$ (Figure \ref{fig: pv_diagrams}) --- and as a region of higher velocity dispersion to the SE of the nucleus in the moment 2 maps (Figures \ref{fig: large_fov_moment_maps} and \ref{fig: moment_maps}) --- is not clear. On one hand, it may represent spatially-extended (220\;\textless\;$r$\;\textless\;440\;pc) AGN-driven outflows. However, since the emission extends well beyond the compact radio source, then these outflows would not be driven solely by the jet --- an additional outflow driving mechanism would be required. Based on the similar PAs of the extended emission and the small-scale radio structure, one possibility is that this gas is accelerated by radiatively-driven winds that are collimated by a circumnuclear torus with an axis similar to that of the compact radio source.  Alternatively, due to the velocities not being much above those expected for rotation (as given by the \textsc{BBarolo model}: Section \ref{section: analysis_and_results: disk}), this extended emission may represent material from the merger settling onto the disk. This latter interpretation is supported by the larger-scale ($r$\;\textgreater1\;kpc) filament-like structures seen to the south and west of the primary nucleus (Figure \ref{fig: large_fov_moment_maps}), which likely represent gas disturbed by the merger: these large-scale structures have similar velocity shifts and dispersions to the intermediate-velocity gas seen near the nucleus.

If the intermediate-velocity emission truly did represent AGN-driven outflows, then its mass outflow rates (22--27\;M$_\odot$\;yr$^{-1}$) are higher than those of the previously detected warm ionised (11.3\;M$_\odot$\;yr$^{-1}$: \citealt{Rose2018})\footnote{The warm ionised mass outflow rates and kinetic powers were calculated following the second method of \citet{Rose2018}, which involves using the $v_\mathrm{05}$ parameter as the outflow velocity to account for projection effects; these estimates are likely to be upper limits.} and neutral HI ($\sim$6\;$M_\odot$\;yr$^{-1}$)\footnote{The neutral HI mass outflow rate has been recalculated using the methodology in Appendix C1 of \citet{Holden2023} with the values presented by \citet{Morganti2013_4c1250}; the kinetic power and coupling efficiency have been recalculated by using these values with Equation \ref{eq: kinetic_power} and assuming $L_\mathrm{bol}=4.8\times10^{45}$\;erg\;s$^{-1}$ \citep{Rose2018}.} phases. Furthermore, they are similar to those recently reported for the cold molecular outflows in a sample of nearby QSO2s (8--16\;M$_\odot$\;yr$^{-1}$: \citealt{RamosAlmeida2022}), which were identified using a similar method to that used here for the intermediate velocity gas. When applying a factor of three to account for assumed spherical outflow geometry, as was done by \citet{Fiore2017}, we find (in agreement with \citealt{RamosAlmeida2022}) that these mass outflow rates (24--81\;M$_\odot$\;yr$^{-1}$) fall significantly below the correlation between mass outflow rate and AGN bolometric luminosity given by \citet{Fiore2017}. However, it is important to note that \citet{Lamperti2022} reported a wide range of mass outflow rates for ULIRGs with similar bolometric luminosities (6\;\textless\;$\dot{M}_\mathrm{flow}$\;\textless\;300\;M$_\odot$\;yr$^{-1}$), indicating that the high mass outflow rates found by \citet{Fiore2017} may represent the most extreme cases, and therefore represent the upper envelope to any relationship between mass outflow rate and AGN bolometric luminosity (see also \citealt{Speranza2023}).

\subsection{The compact, high-velocity nuclear outflow}
\label{section: discussion: nuclear_outflow}

While it is possible that the intermediate-velocity emission does not represent outflowing gas, we highlight that the observed velocities of the redshifted nuclear feature (400\;\textless\;$v$\;\textless\;680\;km\;s$^{-1}$) are much higher than what would be expected due to settled gravitational motions ($v$\;\textless\;300\;km\;s$^{-1}$: the projected velocities seen in the disk). Moreover, the emission of this component is compact ($r$\;\textless\;120\;pc), unresolved in the combined datacube, and on a similar radial scale to both the compact radio structure ($r$\;\textless\;0.06\;arcseconds; $r$\;\textless\;130\;pc: Figure \ref{fig: outflow_moment_map}a) and the other outflow phases ($r$\;\textless\;100\;pc: \citealt{Morganti2013_4c1250, Tadhunter2018}). Therefore, it is much more likely that we are seeing a nuclear outflow, rather than an inflow or settling gas from the merger. Furthermore, the fact that the high-velocity emission is spatially offset from the nucleus (by (3.2$\pm$0.2)$\times10^{-2}$\;arcseconds) along the direction of the small-scale radio structure (PA=$151^\circ$) suggests a relation between the two. This is consistent with acceleration by shocks, which may be induced by either a jet or a broader, radiatively-driven wind that is collimated by the circumnuclear torus. Given that the compact CO(1--0) outflow is seen to be offset along the direction of the radio structure (which would not necessarily be expected in the radiatively-driven scenario), and that the neutral atomic outflow component (detected in HI absorption) is seen to be coincident with the radio hotspot \citep{Morganti2013_4c1250}, we favour the jet-driven interpretation. In this context, it is interesting to note the southern bend in the small-scale radio structure, which may be due to the jet colliding with dense molecular gas, being deflected, and thereby launching the nuclear outflow.

The cold molecular mass outflow rate for this nuclear component ($\dot{M}_\mathrm{flow}$$\sim$230\;M$_\odot$yr$^{-1}$) is more than an order of magnitude higher than those previously derived for the other outflow phases in F13451+1232, lies within the range of values for molecular outflows previously detected in ULIRGs (a few to thousands of solar masses per year: \citealt{Cicone2014, Pereira-Santaella2018, Fluetsch2019, Herrera-Camus2020, Lamperti2022}), and is close to (within 1$\sigma$) of the correlation between cold molecular mass outflow rate and AGN bolometric luminosity found by \citet{Fiore2017} when corrected for assumed spherical outflow geometry.

Overall, the kinetic power of the cold molecular phase in F13451+1232 ($6.6\times10^{43}$\;ergs$^{-1}$) accounts for $\sim$1.4\;per\;cent of L$_\mathrm{bol}$ --- approximately a factor of three higher than the combined kinetic power of the neutral HI ($\sim$0.04\;per\;cent of $L_\mathrm{bol}$: \citealt{Morganti2013_4c1250}) and warm ionised (0.49\;per\;cent of $L_\mathrm{bol}$: \citealt{Rose2018}) outflow phases. Therefore, for all phases, the \textit{total} outflow kinetic power is $\sim$2\;per\;cent of the AGN bolometric luminosity. While this is consistent with the requirements of simulations of outflows launched by AGN in galaxy mergers (0.5\;\textless\;$\epsilon_\mathrm{f}$\;\textless\;10\;per\;cent: e.g. \citealt{DiMatteo2005, Hopkins2010}) --- and thus may be interpreted as the outflows having an impact on the evolution of the galaxy --- we highlight that such comparisons must be done with care, and that it is unclear if interpretations can be made robustly (see \citealt{Harrison2018}).

Finally, we emphasize that even if we assume a CO-to-H$_\mathrm{2}$ conversion factor that is at the lower end of the range of values used in the literature (i.e. $\alpha_\mathrm{CO}=0.3$\;M$_\odot$\;(K\;km\;s$^{-1}$\;pc$^2$)$^{-1}$ for optically thin gas: \citealt{Oosterloo2017, Oosterloo2019}), the derived cold molecular, nuclear mass outflow rate is still much higher than those of the other phases ($\sim$86\;M$_\odot$yr$^{-1}$). Conversely, if a conversion factor that is characteristic of molecular clouds in the Milky Way is assumed ($\alpha_\mathrm{CO}=4.3$\;M$_\odot$\;(K\;km\;s$^{-1}$\;pc$^2$)$^{-1}$: \citealt{Bolatto2013}), the mass and kinetic power of this outflow increase significantly ($\dot{M}_\mathrm{out}\sim1240$\;M$_\odot$\;yr$^{-1}$; $\epsilon_\mathrm{f}\sim7$\;per\;cent).

\section{Conclusions}
\label{section: conclusions}

Through the analysis of high-spatial-resolution CO(1--0) observations of the primary nucleus of the ULIRG F13451+1232, we have detected and characterised a kpc-scale circumnuclear disk and spatially-extended (0.0\;\textless\;$r$\;\textless\;0.3\;arcseconds; 0\;\textless\;$r$\;\textless\;660\;pc), intermediate-velocity emission (300\;\textless\;$|v|$\;\textless\;400\;km\;s$^{-1}$). This intermediate velocity molecular gas may represent AGN radiation-driven outflows, or merger material settling onto the disk. Furthermore, we have also presented evidence for a compact ($r$\;\textless\;120\;pc), cold molecular outflow (400\;\textless\;$v$\;\textless\;680\;km\;s$^{-1}$) at the position of the nucleus with a mass outflow rate of $\dot{M}_\mathrm{out}\sim$230\;M$_\odot$\;yr$^{-1}$ and a kinetic power that is $\sim$1.4\;per\;cent of the AGN bolometric luminosity.

Overall, our detection of cold molecular outflow(s) in F13451+1232 changes the scenario of the AGN-driven outflows in this object: previous studies found that the neutral atomic HI and warm ionised phases had relatively modest mass outflow rates, whereas here we find that the cold molecular outflow is carrying over an order of magnitude more mass (and several times the kinetic power) than the warm ionised and neutral phases. Similar results have been found for other types of active galaxy, where the colder phases have been seen to carry the majority of the outflowing mass \citep{RamosAlmeida2019, Holden2023, HoldenTadhunter2023, Speranza2023}. This further demonstrates the need for multi-wavelength observations to fully quantify the impact of AGN-driven outflows. Moreover, a similarly compact ($r$\;\textless\;120\;pc) but more massive ($\dot{M}_\mathrm{out}\sim$650\;M$_\odot$yr$^{-1}$) outflow was found in the ULIRG PKS\;1549-79 \citep{Oosterloo2019}, which is also both a merger and a luminous radio source. Our results thus provide important evidence that such powerful radio sources accelerate compact molecular outflows, and that these outflows may carry more mass and power than other gas phases.

Crucially, in demonstrating that cold molecular AGN-driven outflows can be compact ($r$\;\textless\;$120$\;pc), our study emphasises the importance of high-spatial resolution observations when robustly quantifying the impact of AGN-driven outflows on their host galaxies.

\section*{Acknowledgements}

We thank the anonymous referee for their feedback, which helped to improve the quality of this manuscript. LRH and CNT acknowledge support from STFC. This paper makes use of the following ALMA data: ADS/JAO.ALMA\#2019.1.01757.S. ALMA is a partnership of ESO (representing its member states), NSF (USA) and NINS (Japan), together with NRC (Canada), MOST and ASIAA (Taiwan), and KASI (Republic of Korea), in cooperation with the Republic of Chile. The Joint ALMA Observatory is operated by ESO, AUI/NRAO and NAOJ. The European VLBI Network is a joint facility of independent European, African, Asian, and North American radio astronomy institutes. Scientific results from data presented in this publication are derived from the following EVN project code(s): GM062. AA and CRA acknowledge the projects ``Feeding and feedback in active galaxies'', with reference PID2019-106027GB-C42, funded by MICINN-AEI/10.13039/501100011033, and ``Quantifying the impact of quasar feedback on galaxy evolution'', with reference EUR2020-112266, funded by MICINN-AEI/10.13039/501100011033 and the European Union NextGenerationEU/PRTR. MPS acknowledges funding support from the Ram\'on y Cajal program of the Spanish Ministerio de Ciencia e Innovaci\'on (RYC2021-033094-I). IL acknowledges support from PID2022-140483NB-C22 funded by AEI 10.13039/501100011033 and BDC 20221289 funded by MCIN by the Recovery, Transformation and Resilience Plan from the Spanish State, and by NextGenerationEU from the European Union through the Recovery and Resilience Facility.

\section*{Data Availability}

The ALMA CO(1--0) data used in this work is publicly available from the ALMA Science Archive (\url{https://almascience.eso.org/aq/}) with Project Code 2019.1.01757.S. The data used to produce the VLBI 1266\;MHz continuum image presented in Figure \ref{fig: outflow_moment_map}a is publicly available from the EVN archive for experiment GM062B (\url{http://archive.jive.nl/scripts/arch.php?exp=GM062B_060605}).



\bibliographystyle{mnras}
\bibliography{precise_outflow_diagnostics} 




\appendix

\section{BBarolo modelling of the gas disk}
\label{appendix: disk_modelling}

In order to identify non-rotational motions in the central few kiloparsecs of the primary nucleus of F13451+1232, we fit the disk-like kinematics seen in our CO(1--0) data (Section \ref{section: analysis_and_results}; Figure \ref{fig: moment_maps}) with a simple disk model using the \textsc{3DFIT} task from the \textsc{BBarolo} tool \citep{DiTeodoro2015}. The \textsc{3DFIT} task fits a series of concentric rings to the data in three spatial dimensions and three velocity dimensions (the rotational velocity, $v_\mathrm{rot}$; the radial velocity, $v_\mathrm{rad}$, and the velocity dispersion $v_\mathrm{disp}$).

Our fitting procedure followed the methodology outlined in \citet{AlonsoHerrero2018}, \citet{DominguezFernandez2020}, and \citet{RamosAlmeida2022}. Throughout, the centre of the disk model was fixed to be the position of the continuum centre (as measured from a two-dimensional Gaussian fit to the continuum image: Section \ref{section: observations_and_data_reduction}). We first performed an initial fit using the \textsc{3DFIT} task in which the radial velocities of the rings were fixed to be 0\;km\;s$^{-1}$ and the scale height of the disk ($z_\mathrm{0}$) was fixed to 0\;pc. The remaining parameters were allowed to vary; initial values for the rotational velocity and velocity dispersion were based on the flux-weighted velocity shift and velocity dispersion maps (Figure \ref{fig: moment_maps}), the initial value of the inclination ($i_\mathrm{initial}=38^\circ$) was based on the measured ratio of the projected disk major and minor axes while assuming a circular disk, and the initial value for the PA was set to that previously estimated from two-dimensional spectroastrometry of CO(2--1) observations of the disk by \citet{Lamperti2022}. However, limits were imposed on the variation of these free parameters --- between successive rings, the rotational velocity was not allowed to change by more than $\pm$50\;km\;s$^{-1}$, nor were the inclination and PA allowed to change by more than $\pm20^\circ$. The fits used uniform weighting and local normalisation. It should be noted that, since the physical size of the modelled rings is much less than the beam size of the observations, the properties of adjacent rings are correlated. However, this does not have a significant effect on the maximum rotational velocity of the disk model (nor does changing the values of the initial parameters), and therefore does not affect the interpretations made in our analysis.

\begin{figure*}
    \centering
    \includegraphics[width=0.8\linewidth]{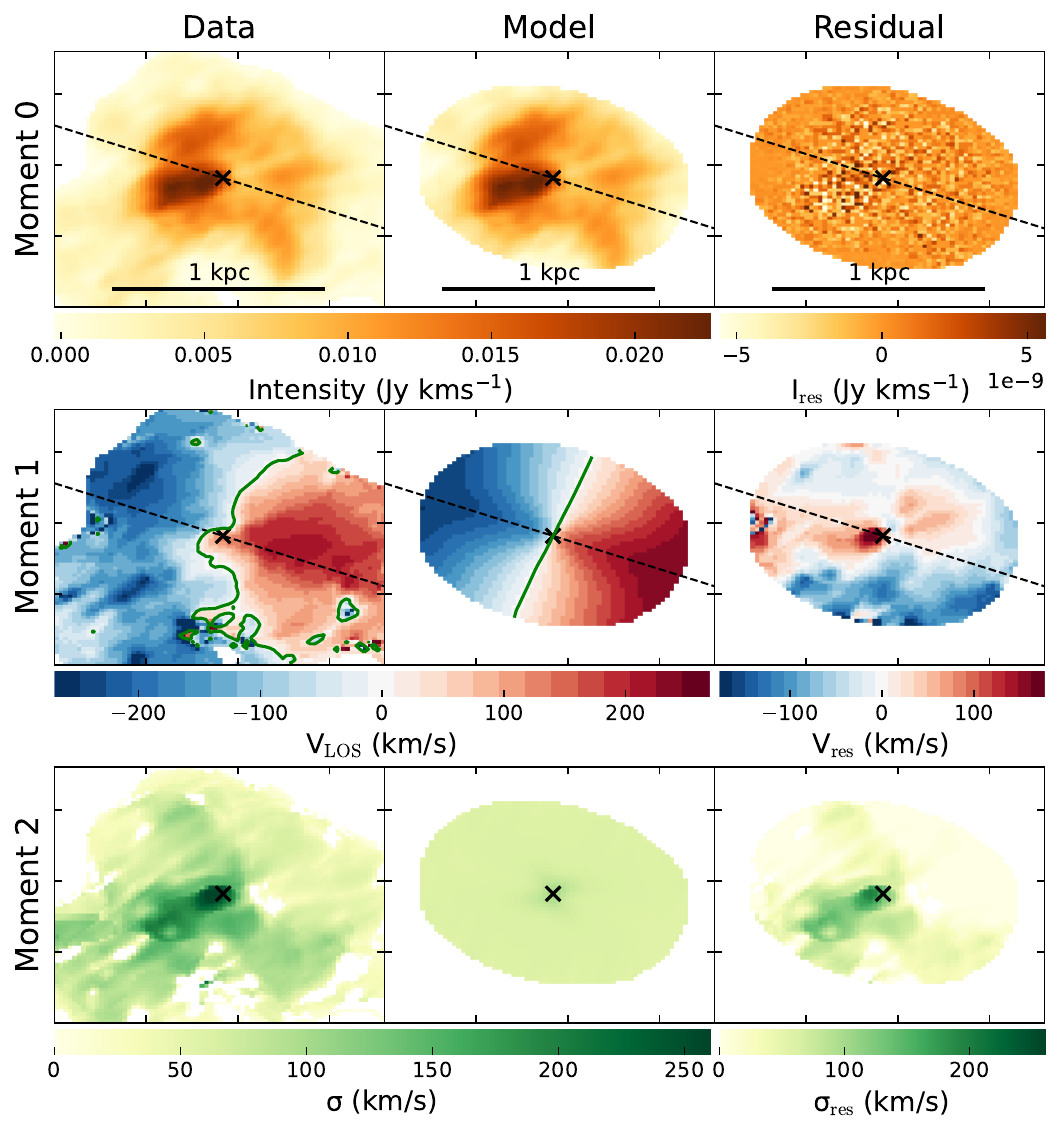}
    \caption{Integrated flux (moment 0; top), flux-weighted velocity (moment 1; middle), and velocity dispersion (moment 2; bottom) maps of the CO(1--0) data (left), \textsc{BBarolo} disk model (middle) and residuals (data - model; right). The ticks have separations of 0.2\;arcseconds. The black dashed line shows the PA of the modelled disk, whereas the black cross shows the disk centre (which was set to be the continuum centre position; see Section \ref{section: observations_and_data_reduction}). The green lines in the middle row panels are isovelocity contours at 0\;km\;s$^{-1}$.}
    \label{fig: bbarolo_model}
\end{figure*}

In order to ensure that the overall radial extent and kinematics of the disk were accurate --- and to alleviate potential problems with fit degeneracy arising from having a high number of free parameters --- we then used the results from the initial fit as a basis for a second fit. For this second run, we fixed the velocity dispersion, inclination, and PA to the values determined from the first run, so that only the rotational velocity was allowed to vary.  We present the final model parameters for each ring in Table \ref{tab: bbarolo_model}; in Figure \ref{fig: bbarolo_model}, we present the moment maps and residuals of the final disk model. High flux-weighted velocities ($\sim$200\;km\;s$^{-1}$) and velocity dispersions ($\sigma\sim$200\;km\;s$^{-1}$) due to non-rotational kinematics can be seen in the residual maps near the centre of the disk.

By taking an aperture around the disk in our CO(1--0) datacube and integrating between $-310$\;\textless\;$v$\;\textless\;$310$\;km\;s$^{-1}$, we find a disk mass of $1.1\times10^{10}$\;M$_\odot$ (using Equation \ref{eq: co_luminosity} and $\alpha_\mathrm{CO}=0.8$\;M$_\odot$\;(K\;km\;s$^{-1}$\;pc$^2$)$^{-1}$). \citet{Lamperti2022} also detected this nuclear disk using CO(2-1) spectroastrometry, and found a similar mass and overall properties to those reported here.

Finally, we highlight that the purpose of this modelling was solely to estimate the maximum CO(1--0) velocities that are explainable as rotational motion, and hence identify outflow kinematics; the detailed parameters of the disk model presented here are not intended to be a robust representation of the disk's properties.

\begin{table*}
    \begin{tabular}{cccccc}
    $r_\mathrm{ring}$\;(arcseconds) & $r_\mathrm{ring}$\;(pc) & v$_\mathrm{rot}$\;(km\;s$^{-1}$) & v$_\mathrm{disp}$\;(km\;s$^{-1}$) & $i$\;$(^\circ)$  & PA\;$(^\circ)$ \\ \hline
    0.013     & 28.5    & 247.969    & 5.524      & 62.218   & 242.005   \\
    0.040     & 87.6    & 249.627    & 24.639     & 58.627   & 236.155   \\
    0.067     & 146.7    & 211.864    & 91.100     & 55.868   & 235.867   \\
    0.094     & 205.8    & 249.245    & 72.507     & 53.826   & 239.504   \\
    0.121     & 264.9    & 267.474    & 65.771     & 52.384   & 245.429   \\
    0.148     & 324.0    & 280.596    & 62.837     & 51.425   & 252.007   \\
    0.175     & 383.1    & 293.392    & 61.210     & 50.834   & 257.600   \\
    0.202     & 442.2    & 311.494    & 68.331     & 50.494   & 260.572   \\
    0.229     & 501.0    & 301.110    & 61.533     & 50.289   & 259.286   \\
    0.256     & 560.4    & 314.645    & 58.314     & 50.102   & 252.106  
    \end{tabular}
    \caption{Final parameters for each ring of the \textsc{BBarolo 3DFit} disk model. The distance of each ring to the fixed disk centre ($r_\mathrm{ring}$) is given in both arcseconds and parsecs. Here, v$_\mathrm{rot}$ and v$_\mathrm{disp}$ are the rotational velocities and velocity dispersions of each ring, respectively.}
    \label{tab: bbarolo_model}
\end{table*}

\clearpage
\section{The CO(1--0) nuclear outflow observed at different spatial resolutions}
\label{appendix: beam_sizes}

\begin{figure}
    \centering
    \includegraphics[width=\linewidth]{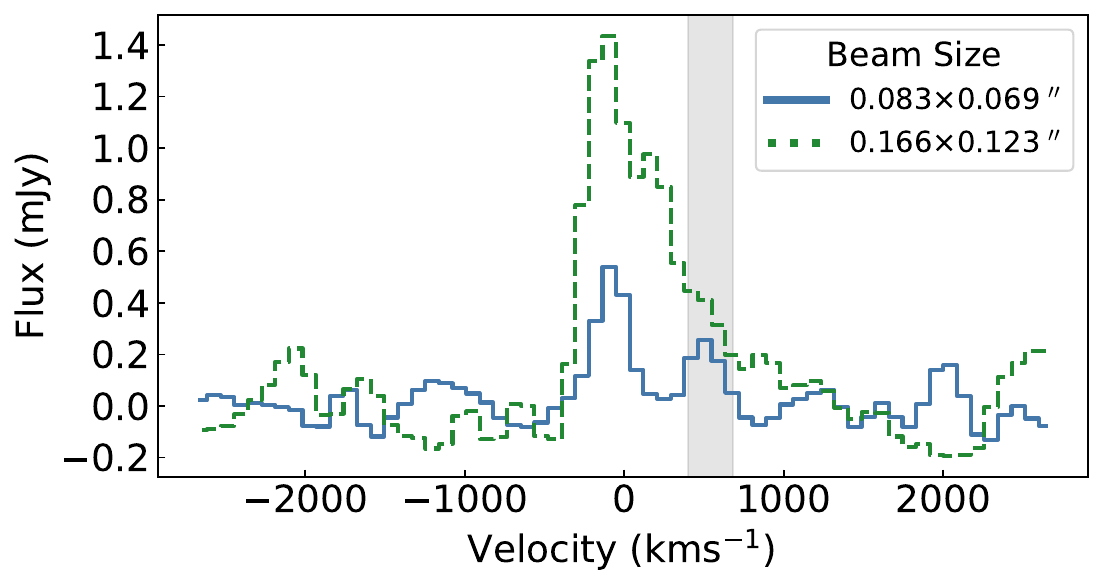}
    \caption{CO(1--0) line profiles of the primary nucleus of F13451+1232 extracted from the two sets of observations of differing array configurations (hence beam sizes) that comprise the datacube presented in our analysis (see Section \ref{section: observations_and_data_reduction}). These profiles were extracted from the respective cubes using apertures matched to the beam sizes. It can be seen that the flux in the high-velocity redshifted component (shaded in grey) is less in the cube with the smaller beam size.}
    \label{fig: beam_comparison}
\end{figure}

The CO(1--0) data that we present in this work was produced from the combined visibilities of two observations, which were taken with different array configurations and thus have different beam sizes ($0.083\times0.069$\;arcseconds and 0.166$\times$0.123\;arcseconds; see Section \ref{section: observations_and_data_reduction}). Here, we discuss the detection of the high-velocity, redshifted, compact feature seen at the position of the nucleus (Section \ref{section: analysis_and_results}; Figures \ref{fig: moment_maps}, \ref{fig: pv_diagrams}, \ref{fig: channel_maps}, and \ref{fig: outflow_moment_map}) in the datacubes produced from the visibilities of these observations separately.

In Figure \ref{fig: beam_comparison}, we present CO(1-0) line spectra extracted from these cubes; the size of the extraction aperture in each case was matched to the beam size. The high-velocity redshifted feature is well detected in both datacubes in the velocity range 400\;\textless\;$v$\;\textless\;680\;km\;s$^{-1}$, while any residual features of comparable flux are not present in both cubes at the same velocity. Moreover, its absolute flux is less in the smaller-beam cube, indicating that the beam is not recovering all of the flux and thus implying that the feature is (partially) spatially resolved. Considering the arguments presented in Section \ref{section: analysis_and_results: potential_effects_of_continuum_subtraction}, we conclude that this feature represents real emission from compact, high-velocity molecular gas near the nucleus.

Since we argue that this molecular gas is a nuclear outflow (Section \ref{section: discussion: nuclear_outflow}), the fact that it is partially spatially resolved by the smaller beam observations allows us to constrain the outflow radius to be in the range 88\;\textless\;$\Delta{R}$\;\textless\;182\;parsecs (0.042\;\textless\;$\Delta{R}$\;\textless\;0.083\;arcseconds; i.e. half of the beam major axes of the two observations). Considering that the feature does not appear to be spatially resolved in the channel maps (Section \ref{section: analysis_and_results: channel_maps}; Figure \ref{fig: channel_maps}) produced from the combined datacube (beam size: $0.113\times0.091$\;arcseconds; see Section \ref{section: observations_and_data_reduction}), this indicates that it is partially spatially-resolved our observations --- we thus take 120\;pc (0.0545\;arcseconds) as an upper limit on the outflow radius.


\bsp	
\label{lastpage}
\end{document}